\pdfoutput=1

\documentclass[11pt,twoside,a4paper,cmspaper,final,collab]{cms-tdr}
\def\svnVersion{175957}\def\cmsCernNo{2012-246}\def\cmsCernDate{\today}\def\cmsMessage{Submitted to the Journal of High Energy Physics}

\begin{document}\cmsNoteHeader{FWD-11-004}

\hyphenation{had-ron-i-za-tion}
\hyphenation{cal-or-i-me-ter}
\hyphenation{de-vices}
\RCS$HeadURL: svn+ssh://svn.cern.ch/reps/tdr2/papers/FWD-11-004/trunk/FWD-11-004.tex $
\RCS$Id: FWD-11-004.tex 150095 2012-10-01 15:38:05Z liwenbo $
\cmsNoteHeader{FWD-11-004}

\newcommand{\e}{\ensuremath{\cmsSymbolFace{e}\xspace}}
\newcommand{\ee}{\ensuremath{\cmsSymbolFace{e}^+\cmsSymbolFace{e}^-}\xspace}
\newcommand{\ep}{\ensuremath{\cmsSymbolFace{e}^+}\xspace}
\newcommand{\en}{\ensuremath{\cmsSymbolFace{e}^-}\xspace}
\newcommand{\gaga}{\ensuremath{\gamma\gamma}\xspace}
\newcommand{\p}{\ensuremath{\cmsSymbolFace{p}}\xspace}
\newcommand{\pbar}{\ensuremath{\overline{\cmsSymbolFace{p}}}\xspace}
\newcommand{\ppbar}{\ensuremath{\cmsSymbolFace{p}\overline{\cmsSymbolFace{p}}}\xspace}
\newcommand{\q}{\ensuremath{\cmsSymbolFace{q}}\xspace}
\newcommand{\qbar}{\ensuremath{\overline{\cmsSymbolFace{q}}}\xspace}
\newcommand{\qqbar}{\ensuremath{\cmsSymbolFace{q}\overline{\cmsSymbolFace{q}}}\xspace}
\newcommand{\ExHuME}{{\textsc{ExHuME}}\xspace}
\newcommand{\LPAIR}{{\textsc{lpair}}\xspace}
\newcommand{\superCHIC}{{\textsc{superCHIC}}\xspace}
\newcommand{\JetSet}{{\textsc{JetSet}}\xspace}
\newcommand{\PHOJET}{{\textsc{phojet}}\xspace}
\newcommand{\MBR}{{\textsc{mbr}}\xspace}
\newcommand{\KMR}{{\textsc{kmr}}\xspace}
\newcommand{\Lund}{{\textsc{Lund}}\xspace}
\newcommand{\theo}{\ensuremath{\,\text{(theo.)}}\xspace}
\providecommand{\re}{\ensuremath{\cmsSymbolFace{e}}}

\title{Search for exclusive or semi-exclusive $\gamma\gamma$ production and observation of exclusive and semi-exclusive \ee production in pp collisions at $\sqrt{s}=7\TeV$}

\date{\today}

\abstract{
    A search for exclusive or semi-exclusive $\gamma\gamma$ production, $\p\p \to \p^{(*)}+\gamma\gamma+\p^{(*)}$
    (where $\p^*$ stands for a diffractively-dissociated proton), and the observation
    of exclusive and semi-exclusive \ee production, $\p\p \to \p^{(*)}+ \ee +\p^{(*)}$, in proton-proton 
    collisions at $\sqrt{s}=7\TeV$, are presented. The analysis is based on a data sample
    corresponding to an integrated luminosity of 36\pbinv recorded by the CMS experiment at the LHC
    at low instantaneous luminosities. Candidate $\gamma\gamma$ or $\ee$ events are selected by
    requiring the presence of two photons or a positron
    and an electron, each with transverse energy $\ET > 5.5\GeV$ and pseudorapidity $|\eta| < 2.5$, and
    no other particles in the region $|\eta|<5.2$.
    No exclusive or semi-exclusive diphoton candidates are found in the data.
    An upper limit on the cross section for the reaction
    $\p\p \to \p^{(*)}+\gamma\gamma+\p^{(*)}$, within the above kinematic selections, 
    is set at 1.18\unit{pb} at 95\% confidence level.
    Seventeen exclusive or semi-exclusive dielectron candidates are observed, with an
    estimated background of $0.85\pm0.28\stat$ events, in agreement with
    the QED-based prediction of $16.3\pm1.3\syst$ events.
}

\hypersetup{%
pdfauthor={CMS Collaboration},%
pdftitle={Search for exclusive or semi-exclusive photon pair production and observation of exclusive and semi-exclusive electron pair production in pp collisions at sqrt(s) = 7 TeV},%
pdfsubject={CMS},%
pdfkeywords={CMS, physics}}

\maketitle

\section{Introduction \label{sec:introduction}}
In central exclusive  (hereafter referred to as
``exclusive'', for brevity) production in pp collisions, $\p\p \rightarrow \p+X+\p$, the colliding protons
emerge intact from the interaction, carrying small transverse
momentum (${\lesssim}2\GeV$), and all the energy transferred from the protons goes into a
color-singlet system at central rapidities. No other particles are produced aside from
the central system, and large rapidity gaps, \ie wide regions of rapidity devoid of particles, 
are present. The three main types of exclusive processes are due to $\gamma\gamma$ interactions
(\eg exclusive \ee or $\Pgmp\Pgmm$ production~\cite{Vermaseren:1982cz}), $\gamma$I$\!$P fusion 
(\eg exclusive $\Upsilon$ production~\cite{Klein:2003vd}) and I$\!$PI$\!$P exchange (\eg exclusive $\gamma\gamma$
or Higgs boson production~\cite{Albrow:2010yb}), where I$\!$P denotes the pomeron, a strongly interacting
color-singlet $t$-channel exchange with the vacuum quantum numbers~\cite{Nachtmann1,Nachtmann2}.

\begin{figure}[hbtp]
\centering
\includegraphics[ width=0.45\textwidth ]{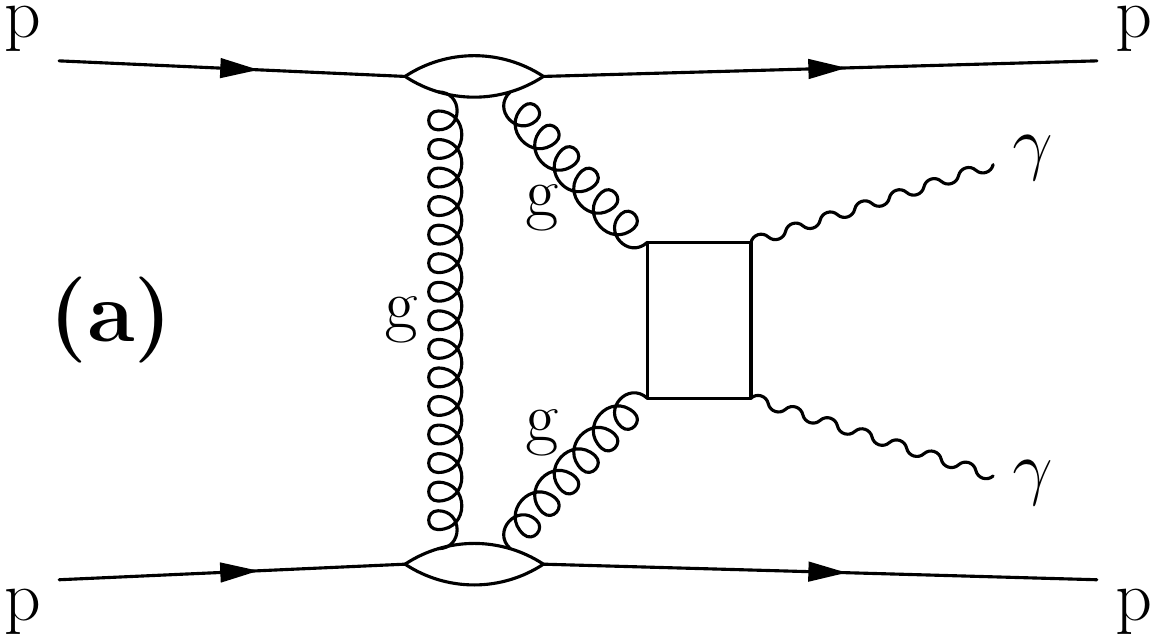} \hspace{2em}
\includegraphics[ width=0.45\textwidth ]{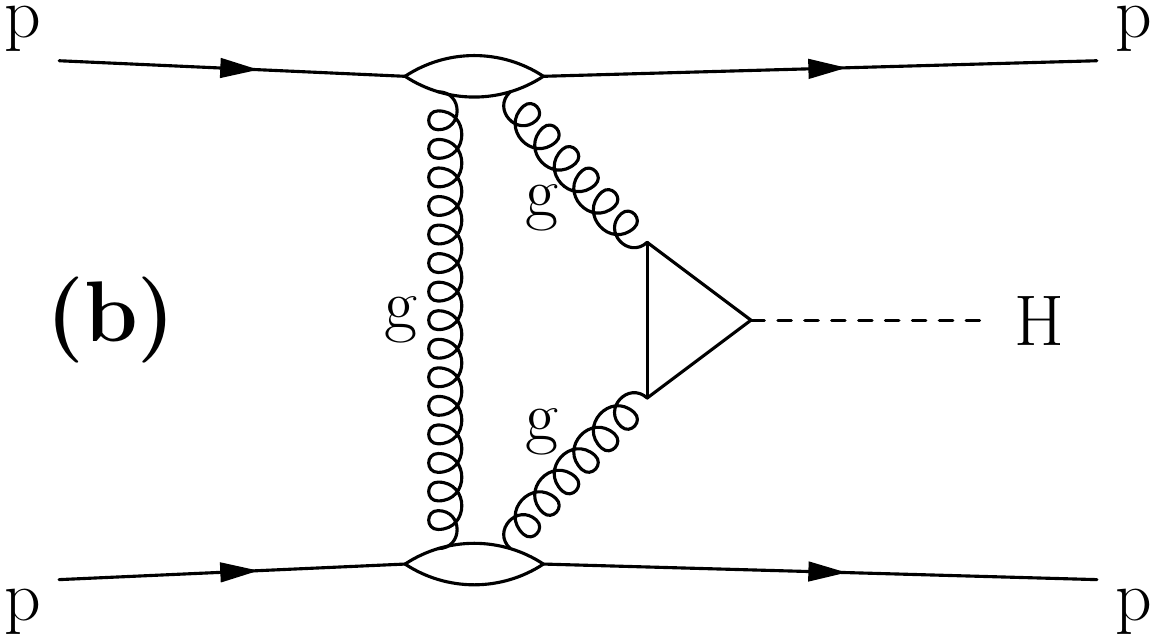}
\caption{The dominant diagrams for (a) exclusive diphoton
production and (b) exclusive Higgs boson production in pp collisions.
Note the screening gluon that cancels the color flow
from the interacting gluons and therefore allows the protons
to stay intact. For exclusive $\gamma\gamma$
production, the contributions from $\qqbar\rightarrow
\gamma\gamma$ and $\gamma\gamma\rightarrow \gamma\gamma$ are both
theoretically estimated to be less than 1\% of
$\Pg\Pg\rightarrow\gamma\gamma$~\cite{Khoze:2004ak}.
\label{fig:feynman}}
\end{figure}
At the Large Hadron Collider (LHC), exclusive $\gamma\gamma$ (hereafter referred to as ``diphoton'')
events can be produced by means of I$\!$PI$\!$P exchange, interpreted in partonic terms as
$\Pg\Pg \rightarrow \gamma\gamma$ via a quark loop, with an additional ``screening" gluon exchanged to
cancel the color of the interacting gluons, as shown in Fig.~\ref{fig:feynman}(a). The quantum chromodynamics
(QCD) calculation of this diagram is difficult
because the screening gluon has low four-momentum-transfer squared,
$Q^2$. Furthermore, additional inelastic interactions between the protons may
produce particles that destroy the rapidity gaps; this effect is taken into account
by introducing the so-called rapidity-gap survival probability~\cite{Alekhin:2005dx}, which is
poorly known theoretically. The study of exclusive diphoton production may shed light on
diffraction and the dynamics of pomeron exchange. In addition, exclusive diphoton production is closely
related to exclusive Higgs boson production (Fig.~\ref{fig:feynman}(b)), where
the Higgs boson is produced via $\Pg\Pg$ fusion dominantly through a top-quark
loop~\cite{Schafer:1990fz,Bialas:1991wj,Albrow:2001fw,Khoze:2002py,Bzdak:2005dv,Petrov:2007kn,Coughlin:2009tr,Cudell:2010cj}.
Since the QCD part of the calculation, from which most theoretical uncertainties originate,
is the same for $\PH$ and $\gamma\gamma$ production,
and only the calculable matrix elements $\Pg\Pg\rightarrow \gamma\gamma$
and $\Pg\Pg\rightarrow \PH$ are different, exclusive $\gamma\gamma$
production provides an excellent test of the theoretical predictions
for exclusive Higgs boson production.

\begin{figure}[hbtp]
\includegraphics[ width=0.33\textwidth ]{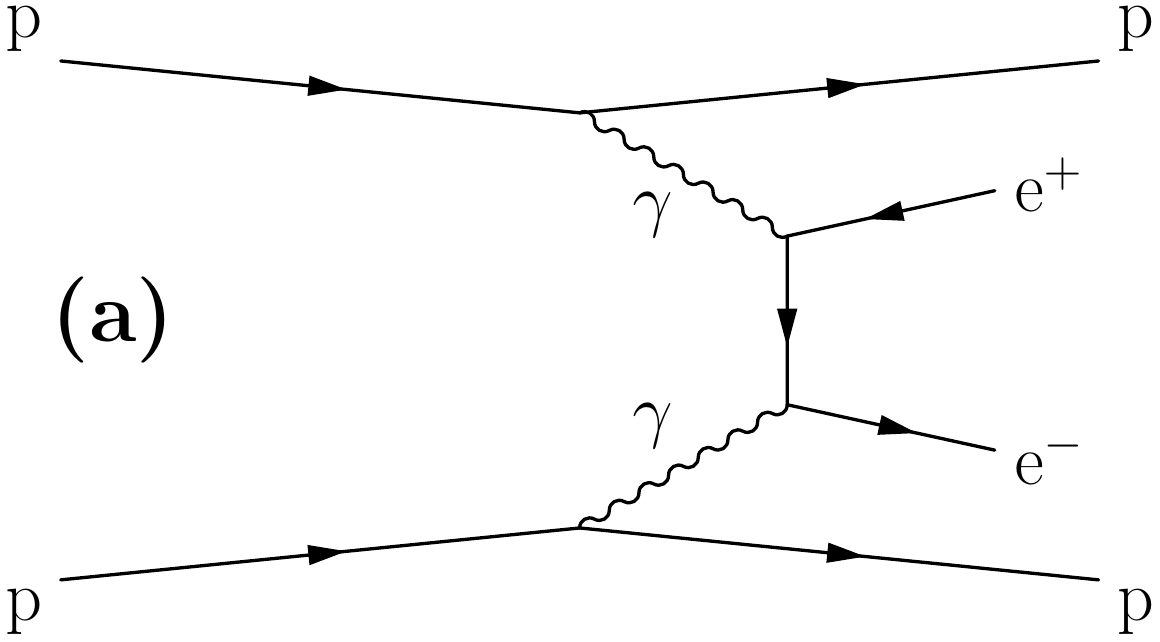}
\includegraphics[ width=0.33\textwidth ]{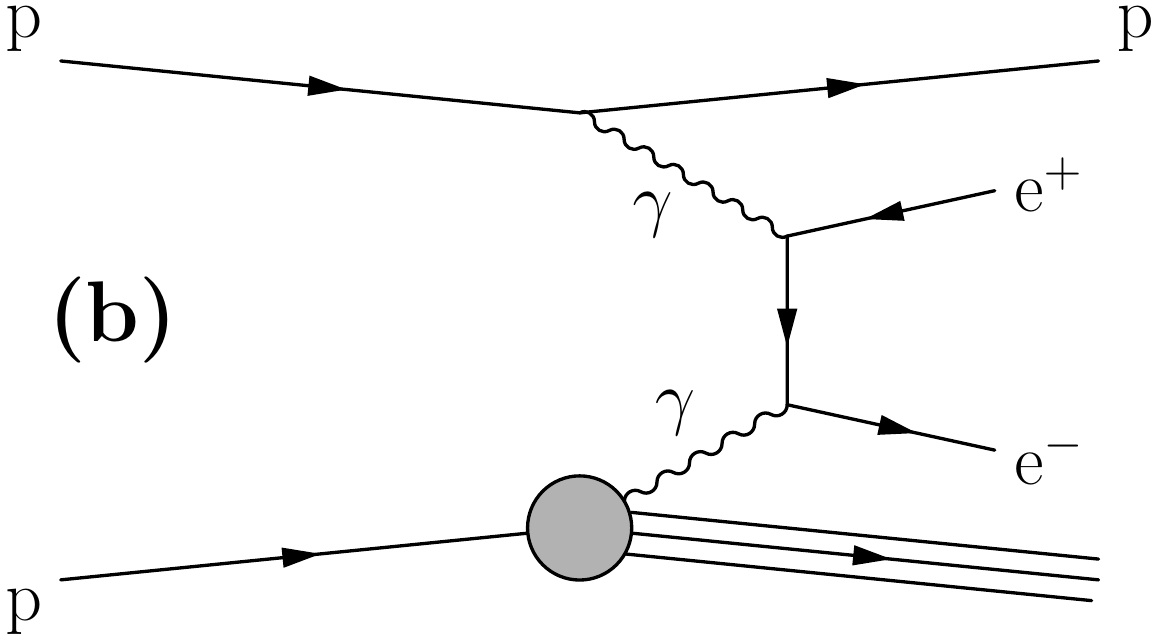}
\includegraphics[ width=0.33\textwidth ]{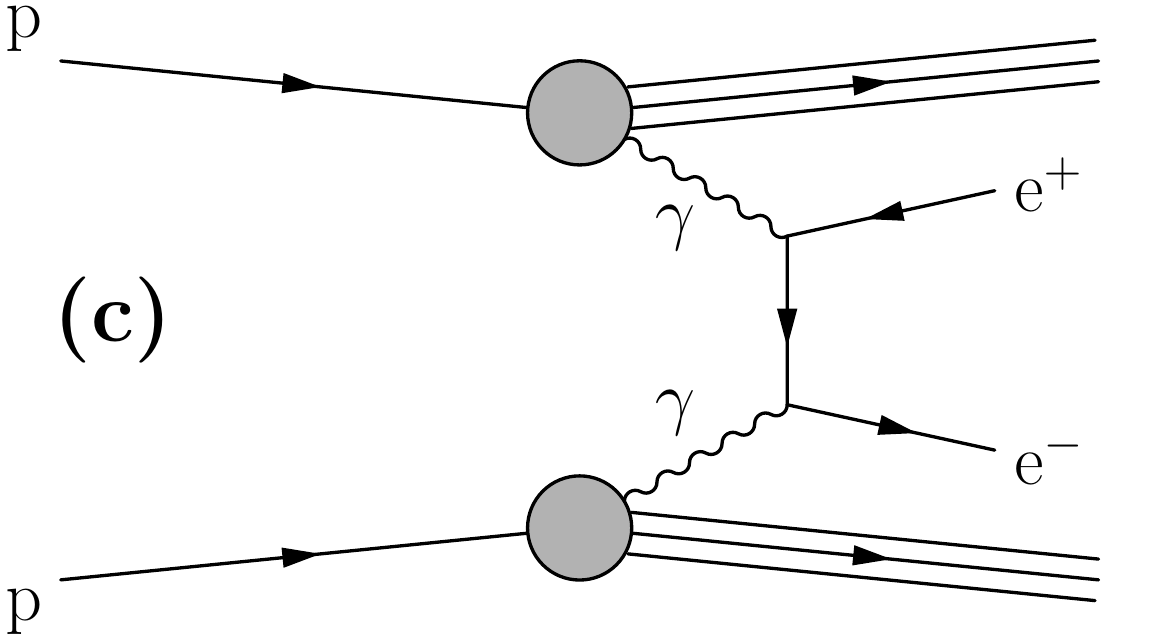}
\caption{The Feynman diagrams for (a) exclusive \ee production and
semi-exclusive \ee production with (b) either or (c) both protons
dissociating in pp collisions. \label{fig:dielectron_feynman}}
\end{figure}
Exclusive \ee (hereafter referred to as ``dielectron'') production via $\gamma\gamma$ interactions is a quantum electrodynamics (QED) process (Fig.~\ref{fig:dielectron_feynman}(a)), and the cross section is
known with an accuracy better than about 1\%; the uncertainty
is dominated by that on the proton electromagnetic form factor~\cite{Bhabha:1935pg,Budnev:1972bb,Budnev1973519}. 
Detailed theoretical studies have shown that in this case the correction due to the rapidity-gap
survival probability is well below 1\% and can be safely neglected~\cite{Khoze:2000db}. 
Exclusive \ee events provide an excellent
control sample for other exclusive processes with less certain theoretical
predictions, such as exclusive $\gamma\gamma$ production.

Semi-exclusive $\gamma\gamma$ and \ee production, involving single- or double-proton
dissociation (Figs.~\ref{fig:dielectron_feynman}(b) and
\ref{fig:dielectron_feynman}(c) for the dielectron case), is also considered as
signal in this analysis, as long as no particles from the proton dissociation have pseudorapidity $|\eta|<5.2$.  
The pseudorapidity $\eta$ is defined as $\eta=-\ln(\tan\frac{\theta}{2})$, where $\theta$ is the polar angle.
This process has larger theoretical uncertainties.
In the rest of this paper, exclusive events
will be referred to as ``el-el" events, while semi-exclusive events with
either or both protons dissociated will be referred to as ``inel-el" and
``inel-inel" events, respectively. The term ``non-exclusive events"
will be used to indicate all other events with
two photons or two electrons and additional activity.

Results on exclusive $\gamma\gamma$ production in \ppbar collisions at a center-of-mass energy of $1.96\TeV$
were obtained by the CDF collaboration~\cite{:2007na,PhysRevLett.108.081801},
and the measured cross sections are consistent with the \KMR~\cite{Khoze:2001xm} predictions. The CDF experiment also
measured the exclusive \ee and $\mu^+\mu^-$ production cross sections~\cite{Abulencia:2006nb,Aaltonen:2009kg,Aaltonen:2009cj}, and the results are in agreement with theory. 
Exclusive $\mu^+\mu^-$ production, which proceeds via the same mechanisms as exclusive \ee production, 
was also measured by the Compact Muon Solenoid (CMS) experiment in pp
collisions at $\sqrt{s}=7 \TeV$~\cite{Chatrchyan:2011ci}, and the result agrees with
the QED-based prediction.

This paper presents a search for exclusive or semi-exclusive $\gamma\gamma$
production, and the observation of exclusive and semi-exclusive \ee production in pp
collisions at $\sqrt{s}=7\TeV$. Since any other
inelastic pp collision occurring in the same bunch crossing as the exclusive interaction
(``pileup'' events) would destroy the rapidity gaps and
make the exclusive interaction unobservable, only a data sample with low pileup contamination is used. 
The data sample was collected in 2010
by the CMS experiment at the LHC, and corresponds to an integrated
luminosity of 36\pbinv. The signal diphoton or dielectron event selection
requires the presence of two photons or two electrons of opposite charge, each with transverse 
energy $\ET > 5.5\GeV$ and pseudorapidity $|\eta| < 2.5$, and no other particles in the
region $|\eta|<5.2$. The two photons or electrons are expected to be
balanced in \ET ($\Delta\ET \sim 0$) and to be back-to-back in azimuthal
angle $\phi$ ($\Delta\phi \sim \pi$), a consequence of the very small
$Q^2$ of the exchanged pomerons or photons.

\section{The CMS detector \label{sec:cmsdetector}}
A detailed description of the CMS detector can be found in
Ref.~\cite{:2008zzk}. The central feature of the CMS apparatus is a
superconducting solenoid of 6\unit{m} internal diameter, providing a
field of 3.8\unit{T}. Within the field volume are the silicon pixel and
strip tracker, the crystal electromagnetic calorimeter (ECAL) and
the brass/scintillator hadron calorimeter (HCAL). Muons are measured
in gas-ionization detectors made by using three technologies: drift tubes (DT), cathode strip
chambers (CSC), and resistive plate chambers. In addition to
the barrel and endcap detectors, CMS has extensive forward
calorimetry. CMS uses a right-handed coordinate system, with the
origin at the nominal interaction point, the $x$ axis pointing to
the center of the LHC ring, the $y$ axis pointing up (perpendicular
to the plane of the LHC ring), and the $z$ axis along the counterclockwise-beam
direction. The polar angle, $\theta$, is measured from the positive
$z$ axis and the azimuthal angle, $\phi$, is measured in the $x$-$y$
plane. The inner tracker measures charged particle trajectories
with transverse momentum \PT from less than
$100\MeV$, and within the pseudorapidity range $\vert\eta\vert<2.5$.
The ECAL provides coverage in the pseudorapidity
range $\vert \eta \vert< 1.479$ in the barrel region (EB) and $1.479 <\vert
\eta \vert < 3.0$ in the two endcap regions (EE). The HCAL provides coverage
for $\vert \eta \vert< 1.3$ in the barrel region (HB) and
$1.3 <\vert \eta \vert < 3.0$ in the two endcap regions (HE). The two
hadronic forward calorimeters (HF) cover the region of $2.9 < |\eta|
< 5.2$. The CMS experiment selects
data by using a two-level trigger system. The first level consists
of custom hardware processors and uses information from the calorimeters and muon
systems. The high-level trigger processor farm further
decreases the event rate before data storage.

\section{Simulation and reconstruction \label{sec:simulation}}
The \ExHuME~1.34 Monte Carlo (MC) event generator~\cite{Monk:2005ji} is used to simulate
exclusive diphoton events and to calculate their
production cross section $\sigma$. The \ExHuME package is an implementation
of the \KMR model~\cite{Khoze:2001xm}. In this model, the two gluons couple
perturbatively to the protons, and produce the $\gamma\gamma$ system through a quark
loop. The calculation includes the Sudakov
factor, which accounts for the probability that no partons are emitted
by the interacting gluons in the evolution up to the hard scale. The
cross section is further suppressed by the rapidity-gap survival probability. A
variety of parton distribution function (PDF) sets have been used, so as to assess
the sensitivity of the cross section calculation to the
low-$x$ gluon density $g(x)$ ($\sigma \sim [g(x)]^4$, where $x$ is the gluon fractional
momentum)~\cite{HarlandLang:2010ep},
which changes significantly in different PDF sets. Semi-exclusive diphoton production
is not well known theoretically, and is not simulated in this analysis.

The \LPAIR~4.0 event generator~\cite{lpair} is used to
simulate both exclusive and semi-exclusive \ee events and to calculate their
production cross sections. For exclusive events, the cross
section depends on the proton electromagnetic form factor. In the case of proton
dissociation, the cross section calculation requires the knowledge of the proton structure
function and the rapidity-gap survival probability. The latter is not included 
in \LPAIR and is taken as 1 in this analysis. In order to
simulate the fragmentation of the excited protons, \LPAIR is interfaced to
the \JetSet~7.408 package~\cite{Sjostrand:1993yb}, where
the \Lund fragmentation model~\cite{Andersson:1983ia} is implemented.

The generated events are further processed through a detailed
simulation of the CMS detector based on
\GEANTfour\cite{Agostinelli:2002hh} and are reconstructed in the same way as the collision data.

Photon candidates are reconstructed~\cite{CMS-PAS-EGM-10-006}
from clusters of ECAL channels around
significant energy deposits, which are merged into so-called
superclusters. The clustering algorithm results in an almost
complete recovery of the energy of photons converting in the
material in front of the ECAL. In the barrel region, superclusters
are formed from 5-crystal-wide strips in $\eta$ centered on the
locally most energetic crystal (seed), and have a variable extension
in $\phi$ (up to $\pm17$ crystals from the seed). In the endcap,
matrices of $5\times5$ crystals (which may partially overlap) around
the most energetic crystals are merged if they lie within a narrow
road in $\eta$ ($\Delta\eta$ = 0.14, $\Delta\phi$ = 0.6\unit{rad}).

The reconstruction of electrons~\cite{Baffioni:2006cd} combines the ECAL and inner-tracker
information. It starts with clusters of energy deposits
in the ECAL, which include the energy due to electron-induced electromagnetic showers
and that of the bremsstrahlung photons emitted along the electron
trajectory. The clusters drive the search for hits in
the pixel detector, which are then used to seed electron
tracks. This is complemented by the usage of the tracker for the
seeding, to improve the reconstruction efficiency at low \PT and in
the transition regions between the ECAL detector elements. Trajectories in
the tracker volume are reconstructed by using a dedicated model of
the electron energy loss, and are fitted with a Gaussian sum filter
(GSF)~\cite{Baffioni:2006cd}. The four-momenta of electrons are
obtained by using the angle from the associated GSF track
and the energy from the combination of the tracker and ECAL information.

\section{Event selection \label{sec:eventselection}}
The selection of signal events proceeds in three steps. Exactly two
photons or two electrons of opposite charge, each with $\ET>5.5\GeV$
and $|\eta|<2.5$, are required to be present in the triggered events.
Then, the events are required to
satisfy the cosmic-ray rejection criteria. Finally, the exclusivity
selection is performed, based on the information from the tracker,
the electromagnetic calorimeter, the hadron calorimeter,
and the muon chambers; this selection requires no additional particles
reconstructed in these subdetectors, and thus suppresses the contribution from
semi-exclusive events and rejects non-exclusive events as well as pileup events.

\subsection{Photon and electron selection \label{sec:photonselection}}
Both diphoton and dielectron candidate events were selected online
by two different triggers corresponding to two subsequent data acquisition periods. Both triggers
required the presence of two electromagnetic showers with
$\ET>5\GeV$. In the second data acquisition period with higher instantaneous luminosities,
the two showers were also required to be separated
in azimuthal angle by at least 2.5\unit{rad}, and a low-activity requirement of less than 10 hadronic 
towers with energy above 5\GeV and $|\eta| < 5.2$ was applied.

The first offline selection step is to require the presence of exactly
two photon candidates or two electron candidates of opposite charge, 
each with $\ET > 5.5\GeV$ and $|\eta| < 2.5$, for the
diphoton and the dielectron analyses, respectively. These photon
or electron candidates are subsequently required to satisfy the identification
criteria described below.

For photons, the energy detected in the HCAL behind the photon
cluster is required to be less than 2\% of the ECAL energy, and the
ECAL cluster-shape parameter~\cite{CMS-PAS-EGM-10-006} is
required to be consistent with that of a photon. The photons are required to be isolated from
other activity in the detector. The isolation parameter is
defined as the scalar sum of the transverse energies of tracks or
calorimeter deposits within $\Delta R = \sqrt{(\Delta \eta)^2
+(\Delta \phi)^2}=0.4$ of the direction of the photon, after excluding
the contribution from the candidate itself. The isolation parameter is required to be less
than $0.001\times \ET + 1.0 \GeV$, $0.006\times \ET +2.5 \GeV$, and
$0.0025\times \ET +2.0 \GeV$ for the tracker, ECAL, and HCAL,
respectively, where \ET is the photon transverse energy in \GeV. The
absence of any hit patterns in the pixel tracker consistent
with those of an electron track is also required in order to discriminate
photons from electrons. No explicit attempt is made to distinguish between
photons and neutral pions when the showers of the two decay photons merge.

For electrons, the same requirements on the HCAL energy and the
cluster shape are applied as in the photon case. The ratio between
the isolation parameter described above (but with $\Delta R = 0.3$) and the
electron \PT is required to be less than $0.05$, $0.3$, and
$0.2$~(barrel) or $0.1$~(endcap), for the tracker, ECAL, and HCAL,
respectively. The difference between the azimuthal angle of the cluster and that of the
direction of the electron track at its vertex is required to be less
than 0.3\unit{rad}; the corresponding difference in pseudorapidity is required to be less than 0.02~(EB)
or 0.03~(EE). The number of missing hits
in front of the first valid hit of the electron track is required
to be ${\le}1$ in order to reject electrons from photon conversions.

\subsection{Cosmic-ray rejection \label{sec:cosmicselection}}
In order to remove cosmic-ray events, the timing of the two photons or electrons,
as measured by the ECAL, is required to be consistent with
that of particles originating from a collision, \ie
$|\text{t}_1|<2\unit{ns}$, $|\text{t}_2|<2\unit{ns}$, and $|\text{t}_1-\text{t}_2|<2\unit{ns}$,
where $\text{t}_i$ is the timing of the $i$-th photon or electron.
Furthermore, the two photon or electron candidates are
required to be separated by more than 2.5\unit{rad} in $\phi$, in order to reject the remaining cosmic-ray 
events in which the cosmic ray is far away from the interaction point in the $x$-$y$ plane.

\subsection{Exclusivity selection \label{sec:exclusivityselection}}
Exclusivity selection criteria are designed to reject events with
particles in the range $|\eta|<5.2$ not associated with the two
photon or electron candidates. More specifically, it is required
that there should be no additional
tracks in the tracker, no additional towers above the noise thresholds in the
calorimeters (EB, EE, HB, HE, and HF), and no track segments in the DTs and CSCs.
An additional track is defined as any track outside a region of
$\Delta \eta < 0.15 \text{ and } \Delta \phi < 0.7$\,rad of the
photons or the electrons. An additional tower in the EB is defined as a tower
above the noise threshold and outside a region of $\Delta \eta <
0.15 \text{ and } \Delta \phi < 0.7$\,rad of the photons or the electrons,
while in the EE the region is $\Delta \eta < 0.15 \text{ and }
\Delta \phi < 0.4$\,rad. An additional tower in the HB, HE, and HF is
defined as any tower above the noise thresholds. The noise thresholds
are determined from non-interaction events. The values of the noise
thresholds are 0.52\GeV, 2.18\GeV, 1.18\GeV, 1.95\GeV, and 9.0\GeV
for the EB, EE, HB, HE, and HF, respectively, and are applied in
energy rather than \ET.

The numbers of diphoton and dielectron candidates in the data sample
remaining after each selection step are listed in
Table~\ref{tab:diphotoncandidatenumber}.
\begin{table}[hbtp]
\centering \topcaption{Numbers of diphoton and dielectron candidates
remaining after each selection step.
\label{tab:diphotoncandidatenumber}}
\begin{tabular}{l|r||l|r}
\hline \multicolumn{2}{c||}{Diphoton analysis} &
\multicolumn{2}{c}{Dielectron analysis}    \\
\hline Selection criterion     & Events remaining      & Selection criterion     & Events remaining \\
\hline Trigger                 & 3\,023\,496           & Trigger                 & 3\,023\,496 \\
\hline Photon reconstruction   & 1\,683\,526           & Electron reconstruction & 132\,271 \\
\hline Photon identification   & 40\,692               & Electron identification & 1\,668 \\
\hline Cosmic-ray rejection    & 34\,234               & Cosmic-ray rejection    & 1\,321 \\
\hline Exclusivity requirement & 0                     & Exclusivity requirement & 17 \\
\hline
\end{tabular}
\end{table}

\section{Efficiencies \label{sec:efficiency}}
The overall selection efficiency $\varepsilon$ is defined as
$\varepsilon=\varepsilon_{\gamma \gamma(\ee)}\cdot\varepsilon_{\text{cos}}\cdot\varepsilon_{\text{fsr}}\cdot\varepsilon_{\text{exc}}$, where
$\varepsilon_{\gamma \gamma(\ee)}$ is the efficiency for identifying the two photons or electrons;
$\varepsilon_{\text{cos}}$ is the efficiency for a signal event to pass the cosmic-ray rejection criteria;
$\varepsilon_{\text{fsr}}$ is the probability for a signal event not to be rejected by the exclusivity selection
criteria because of final-state radiation;
and $\varepsilon_{\text{exc}}$ is the probability for a signal event not to be rejected by the exclusivity selection
criteria because of pileup, calorimeter noise, or beam background.

\subsection{Photon and electron efficiency \label{sec:photonefficiency}}
The diphoton efficiency $\varepsilon_{\gamma \gamma}$ is split
into three parts: the reconstruction efficiency
$\varepsilon_{\text{reco}}$, the identification efficiency
$\varepsilon_{\text{id}}$, and the trigger efficiency
$\varepsilon_{\text{trig}}$, \ie \( \varepsilon_{\gamma \gamma} =
\varepsilon_{\gamma \gamma, \text{\,reco}} \cdot
\varepsilon^{2}_{\gamma, \text{\,id}} \cdot \varepsilon_{\gamma
\gamma, \text{\,trig}}\). The reconstruction and trigger
efficiencies are both denoted by the subscript ``$\gamma \gamma$",
rather than just ``$\gamma$", to reflect the fact that these
efficiencies must be calculated per event, rather than per photon,
due to the strong \ET and $\phi$ correlations between the two
photons (balanced in \ET and back-to-back in $\phi$). All these
efficiencies are calculated by using signal MC samples.
The systematic uncertainty of the reconstruction efficiency is
evaluated by shifting the \ET threshold by $\pm$5\%, motivated by the energy scale
uncertainty for low-\ET photons and electrons. The
systematic uncertainty of the identification efficiency is evaluated
by shifting the thresholds of the identification parameters by $\pm$10\%.
The systematic uncertainty of the trigger efficiency is estimated from the
difference of the single-photon trigger efficiency calculated from
interaction (minimum-bias) events in the data and in the MC samples.
A summary of the photon efficiencies for exclusive diphoton events
is listed in Table~\ref{tab:photonefficiency}.

For the dielectron analysis, the same procedure as in the diphoton analysis is used to determine
the electron efficiencies and the
corresponding systematic uncertainties. The results are listed in
Table~\ref{tab:photonefficiency} for both exclusive and
semi-exclusive \ee events.
\begin{table}[hbtp]
\centering \topcaption{Summary of the photon and electron efficiencies
with systematic uncertainties. \label{tab:photonefficiency}}
\begin{tabular}{l|l||l|l|l|l}
\hline \multicolumn{2}{c||}{\raisebox{-0.6em}[0pt][0pt]{Diphoton analysis}} &
\multicolumn{4}{c}{Dielectron analysis} \\
\cline{3-6} \multicolumn{2}{l||}{} & & el-el & inel-el & inel-inel \\
\hline $\varepsilon_{\gamma \gamma, \text{\,reco}}$  & 0.724$\pm$0.087 & $\varepsilon_{\ee, \text{\,reco}}$ & 0.606$\pm$0.055 & 0.663$\pm$0.050 & 0.683$\pm$0.045 \\
\hline $\varepsilon_{\gamma, \text{\,id}}$           & 0.941$\pm$0.003 & $\varepsilon_{\text{e}, \text{\,id}}$   & 0.967$\pm$0.005 & 0.966$\pm$0.005 & 0.960$\pm$0.005 \\
\hline $\varepsilon_{\gamma \gamma, \text{\,trig}}$  & 0.757$\pm$0.050 & $\varepsilon_{\ee, \text{\,trig}}$ & 0.655$\pm$0.024 & 0.708$\pm$0.018 & 0.683$\pm$0.013 \\
\hline \hline $\varepsilon_{\gamma \gamma}$          & 0.485$\pm$0.067 & $\varepsilon_{\ee}$                & 0.371$\pm$0.037 & 0.438$\pm$0.035 & 0.430$\pm$0.030 \\
\hline
\end{tabular}
\end{table}

\subsection{Cosmic-ray rejection efficiency \label{sec:cosmicefficiency}}
For exclusive \gaga and \ee events, since the efficiency for the requirement of $\Delta \phi > 2.5$\unit{rad} is
100\%, the cosmic-ray rejection efficiency $\varepsilon_{\text{cos}}$ is equal to the efficiency for the timing requirements mentioned in Section~\ref{sec:cosmicselection}. This efficiency is determined by applying the timing requirements to a data sample of $\JPsi\to\ee$ events with invariant mass $3.0< M(\ee) < 3.2\GeV$, which has a negligible
cosmic-ray contamination. This yields $\varepsilon_{\text{cos}}=0.979\pm0.009$ for exclusive \gaga and \ee events.
The quoted systematic uncertainty is
evaluated by shifting the thresholds of the timing requirements by $\pm$5\%, motivated by the
uncertainty of the timing measurement of less than 100\unit{ps}. For semi-exclusive \ee events, the
efficiency for the $\Delta \phi$ requirement is determined from MC to be
0.858 and 0.701 for inel-el and inel-inel events, respectively. A correction factor of 0.979 and 0.932
is subsequently applied for inel-el and inel-inel \ee events in order to take into account the $\Delta \phi$
requirement at the trigger level. The cosmic-ray rejection efficiency for inel-el and
inel-inel \ee events is then estimated to be $0.822\pm0.008$
and $0.639\pm0.006$, respectively.

\subsection{Final-state-radiation efficiency \label{sec:fsrefficiency}}
As a consequence of the exclusivity requirements, signal diphoton events with
either or both photons converting into \ee pairs, as well as events that
produce electrons in the tracker detector by Compton scattering,
are vetoed if there are energy deposits above the noise thresholds outside the regions defined in Section~\ref{sec:exclusivityselection}.
The corresponding efficiency
is the final-state-radiation efficiency $\varepsilon_{\text{fsr}}$, and is estimated 
by applying the exclusivity selection
criteria to simulated signal events. The systematic uncertainty is
evaluated by shifting the noise thresholds of the exclusivity
selection criteria by the energy scale uncertainty for
each subdetector. The uncertainty due to the tracker-material budget is negligible and is evaluated by
using a set of realistic tracker-material modifications~\cite{Migliore:1278158} in the simulation.

Likewise, for both exclusive and semi-exclusive dielectron production, if a
final-state electron emits a high-energy bremsstrahlung photon, the
event is vetoed by the exclusivity selection criteria. For the semi-exclusive case, the probability
that a semi-exclusive event is not vetoed because of the particles from the proton dissociation
is also folded into this efficiency, which results in a much lower final-state-radiation
efficiency than for the exclusive case. The same procedure as in the diphoton analysis is used to determine the efficiencies and the uncertainties due to the energy scale. For the semi-exclusive case, the 
additional uncertainty coming from the proton fragmentation model is dominant, and 
is evaluated by using different programs
to simulate the dissociation of the excited protons. The programs considered are \PHOJET~1.12~\cite{Bopp:1998rc,Engel:1995sb}, \PYTHIA~6.422~\cite{Sjostrand:2006za}, \PYTHIA~8.142~\cite{Sjostrand:2007gs}, and \PYTHIA~8.165 with \MBR~\cite{Ciesielski:2012mc}.

\subsection{Exclusivity efficiency \label{sec:exclusivityefficiency}}
\begin{figure}[hbtp]
\centering
\includegraphics[ width=0.5\textwidth ]{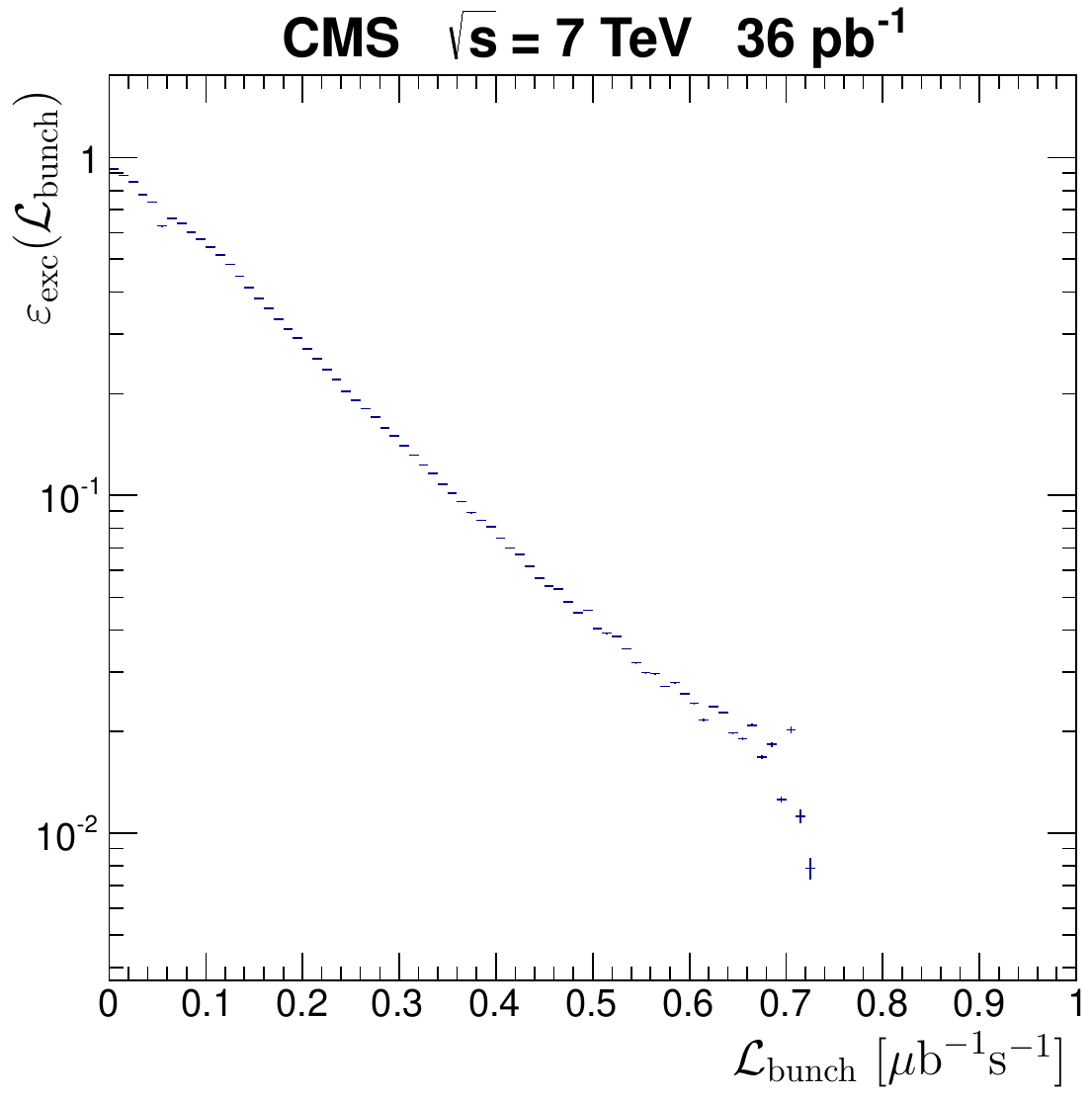}
\caption{Exclusivity efficiency as a function of the bunch-by-bunch
luminosity. \label{fig:exclusivityefficiency}}
\end{figure}
The exclusivity efficiency is the probability that a signal event is not rejected by the exclusivity
selection criteria because of pileup, calorimeter noise, or beam background in the same
bunch crossing, and is determined by using zero-bias events. Zero-bias events are those triggered
solely on the bunch-crossing time. Since
the number of inelastic proton-proton interactions in a given bunch
crossing follows a Poisson distribution and the exclusivity
efficiency is approximately equal to the probability of having no
inelastic collision, the exclusivity efficiency is an
exponential function of the bunch-by-bunch instantaneous luminosity:

\vspace{-1em}
\[\varepsilon_{\text{exc}}(\Lumi_{\text{bunch}}) = \frac{N_{\text{zero-bias}}^{\text{exc}}(\Lumi_{\text{bunch}})}{N_{\text{zero-bias}}(\Lumi_{\text{bunch}})} \approx \re^{-\overline{n}} =
\re^{-\Lumi_{\text{bunch}}\cdot \sigma_{\text{inelastic}} /f}\] where
$N_{\text{zero-bias}}^{\text{(exc)}}$ is the number of zero-bias
events with (exc) or without the exclusivity requirements,
$\overline{n}$ is the average number of inelastic interactions per
bunch crossing for a given bunch-by-bunch luminosity
$\Lumi_{\text{bunch}}$, and $f = 11\,246$\unit{Hz} is the LHC revolution
frequency. The exclusivity efficiency is shown in Fig.~\ref{fig:exclusivityefficiency} as a function of the
bunch-by-bunch luminosity, calculated with a zero-bias data sample taken during the same
data acquisition period as that of the signal sample.

The average exclusivity efficiency is calculated by using the
following equation~\cite{Abulencia:2006nb}:

\vspace{-1em}
\[\varepsilon_{\text{exc}} = \frac{\int{\frac{\text{d}N_{\text{zero-bias}}}{\text{d}\Lumi_{\text{bunch}}} \cdot \Lumi_{\text{bunch}} \cdot \varepsilon_{\text{exc}}(\Lumi_{\text{bunch}}) \cdot \text{d}\Lumi_{\text{bunch}}}}{\int{\frac{\text{d}N_{\text{zero-bias}}}{\text{d}\Lumi_{\text{bunch}}} \cdot \Lumi_{\text{bunch}} \cdot
\text{d}\Lumi_{\text{bunch}}}} \] where the weight
$\Lumi_{\text{bunch}}$ in the integrations reflects the fact that
the probability of a process taking place in a given bunch crossing
is proportional to the corresponding bunch-by-bunch
luminosity. The average exclusivity efficiency is
\(\varepsilon_{\text{exc}} = 0.145 \pm 0.008 \), where the
uncertainty is evaluated by varying the noise
thresholds of the exclusivity selection criteria by $\pm$5\%. This
efficiency is dominated by the losses due to pileup.

Table~\ref{tab:efficiencysummary} lists a summary of the
efficiencies for both the diphoton and the dielectron analyses.
\begin{table}[hbtp]
\centering \topcaption{Summary of the efficiencies for both the
diphoton and the dielectron analyses. The quoted uncertainties are systematic.
\label{tab:efficiencysummary}}
\begin{tabular}{l|l||l|l|l|l}
\hline
\multicolumn{2}{l||}{\raisebox{-0.6em}[\baselineskip][0pt]{Diphoton analysis}} &
\multicolumn{4}{c}{Dielectron analysis} \\
\cline{3-6} \multicolumn{2}{l||}{} & & el-el & inel-el & inel-inel \\
\hline $\varepsilon_{\gamma \gamma}$ &  0.485 $\pm$ 0.067      & $\varepsilon_{\ee}$                  & 0.371 $\pm$ 0.037 & 0.438 $\pm$ 0.035 & 0.430 $\pm$ 0.030 \\
\hline $\varepsilon_{\text{cos}}$    &  0.979 $\pm$ 0.009      & $\varepsilon_{\text{cos}}$           & 0.979 $\pm$ 0.009 & 0.822 $\pm$ 0.008 & 0.639 $\pm$ 0.006 \\
\hline $\varepsilon_{\text{fsr}}$    &  0.972 $\pm$ 0.005      & $\varepsilon_{\text{fsr}}$           & 0.927 $\pm$ 0.005 & 0.666 $\pm$ 0.049 & 0.299 $\pm$ 0.041 \\
\hline $\varepsilon_{\text{exc}}$    &  0.143 $\pm$ 0.008      & $\varepsilon_{\text{exc}}$           & 0.143 $\pm$ 0.008 & 0.143 $\pm$ 0.008 & 0.143 $\pm$ 0.008 \\
\hline \hline $\varepsilon$          &  0.0660 $\pm$ 0.0099    & $\varepsilon$                        & 0.0481$\pm$0.0055 & 0.0343$\pm$0.0042 & 0.0117 $\pm$0.0019 \\
\hline
\end{tabular}
\end{table}

\section{Backgrounds}
For diphoton production, the following background
processes are considered: non-exclusive events, exclusive \ee production, cosmic-ray
events, and exclusive $\pi^0\pi^0$ production.

The non-exclusive background consists of non-exclusive events with
particles passing through the cracks between the calorimeter
elements, or with energy deposits below the noise thresholds, so that they appear exclusive. 
In order to estimate the amount of
this background, the two-dimensional
distribution of the numbers of additional tracks and additional
towers for diphoton events, with all selection
criteria applied except the exclusivity requirements, is fitted and
then extrapolated to the signal region, \ie the bin with no additional tracks
or towers. This yields a non-exclusive background of $1.68\pm0.40$ events.

Exclusive \ee events can be misidentified as diphoton events if
neither electron track is reconstructed or both electrons undergo
hard bremsstrahlung. This contribution is estimated by assuming a
single-electron misidentification probability of 8\%, as
determined from simulated exclusive \ee events, for the 17
\ee candidates found in the data (Table~\ref{tab:diphotoncandidatenumber}), 
which results in a background of $0.11 \pm 0.03$ events.

The background from cosmic-ray events is
evaluated by measuring the density of cosmic-ray events outside the
signal region described in Section~\ref{sec:cosmicselection} and then extrapolating that density into
the signal region. This results in a probability of 0.46\% that a
diphoton candidate is due to a cosmic ray.

Exclusive $\pi^0\pi^0$ production ($\pi^0\rightarrow\gamma\gamma$)~\cite{HarlandLang:2011qd}
can be a background to diphoton production if the two pions are both
misidentified as photons. A simulation carried out with the \superCHIC~1.41 event 
generator~\cite{HarlandLang:2009qe} is used to calculate the
cross section and derive the selection efficiency. Fewer than $10^{-4}$
exclusive diphoton candidates are expected to originate from $\pi^0\pi^0$ events.
Therefore, the background from exclusive $\pi^0\pi^0$
production, even with conservative theoretical uncertainties, is
negligible. The background from exclusive pair production of other mesons,
\eg $\p\p\to\p+\eta\eta+\p$ ($\eta\rightarrow\gamma\gamma$), is also estimated to be negligible 
because of the low production cross sections (which are similar to that of exclusive $\pi^0\pi^0$ production).
Exclusive $\gamma\pi^0$ or $\gamma\eta$ production is forbidden by C-parity conservation. 
Exclusive single-meson production, \eg $\p\p\to\p+\eta+\p\to\p+\gamma\gamma+\p$, is completely 
removed by the requirement $\ET(\gamma)>5.5\GeV$, complemented by $\Delta\phi(\gamma\gamma)>2.5\unit{rad}$, 
which selects events with $M(\gamma\gamma)\gtrsim11\GeV$.

For dielectron production, the following background
processes are considered: non-exclusive events, exclusive $\Upsilon$ production,
cosmic-ray events, and exclusive $\pi^+\pi^-$
production.

The non-exclusive background is estimated by using the distribution of the numbers of additional tracks and additional
towers for dielectron events with all selection
criteria applied except the exclusivity requirements, after subtracting the contributions
from both exclusive and semi-exclusive \ee production expected from the simulation.
This background is estimated to be of $0.80\pm 0.28$ events.

The background from exclusive $\Upsilon$ production via $\gamma$I$\!$P fusion
($\gamma$I$\!$P$\rightarrow \Upsilon(\text{1S,2S,3S})
\rightarrow \ee$)~\cite{Klein:2003vd} is completely removed by the $\ET> 5.5\GeV$
requirement on the electrons, which corresponds to
$M(\ee)\gtrsim11\GeV$, well above the $\Upsilon(\text{3S})$
mass ($10.36\GeV$) even taking into
account the \ee mass resolution of $\sim$150\MeV.

The cosmic-ray background contamination, estimated with the same method as for the
diphoton analysis, is 0.3\%, \ie $0.05\pm0.01$ events.

Exclusive $\pi^+\pi^-$ production via I$\!$PI$\!$P exchange~\cite{HarlandLang:2011qd} can be a background
to \ee production if the two pions are both misidentified as electrons.
The cross section, calculated with \superCHIC, is less
than 0.1\% of that for exclusive \ee production, which translates into a
negligible background. This is consistent with the fact that no additional
candidates are found, after removing the requirement of no HCAL energy behind the
electron shower (a high-energy deposit in the HCAL is the signature of a pion).

A summary of the background processes for both the diphoton
and the dielectron analyses is listed in Table~\ref{tab:background}.
The non-exclusive background is the largest contribution in both analyses.
\begin{table}[hbtp]
\centering \topcaption{Background event yields expected for both the
diphoton and the dielectron analyses. The quoted uncertainties are statistical.%
\label{tab:background}}
\begin{tabular}{l|l||l|l}
\hline \multicolumn{2}{c||}{Diphoton analysis} &
\multicolumn{2}{c}{Dielectron analysis} \\
\hline Background             & Events              & Background                                     & Events         \\
\hline Non-exclusive          & $1.68\pm 0.40$      & Non-exclusive                                  & $0.80\pm 0.28$ \\
\hline Exclusive \ee          & $0.11\pm 0.03$      & Exclusive $\Upsilon\text{(1S,2S,3S)}\to\ee$    & Negligible     \\
\hline Cosmic ray             & Negligible          & Cosmic ray                                     & $0.05\pm 0.01$ \\
\hline Exclusive $\pi^0\pi^0$ & Negligible          & Exclusive $\pi^+\pi^-$                         & Negligible     \\
\hline \hline Total           & $1.79\pm 0.40$      & Total                                          & $0.85\pm 0.28$ \\
\hline
\end{tabular}
\end{table}

\section{Results \label{sec:result}}
\begin{figure}[hbtp]
\centering
\includegraphics[ width=0.5\textwidth ]{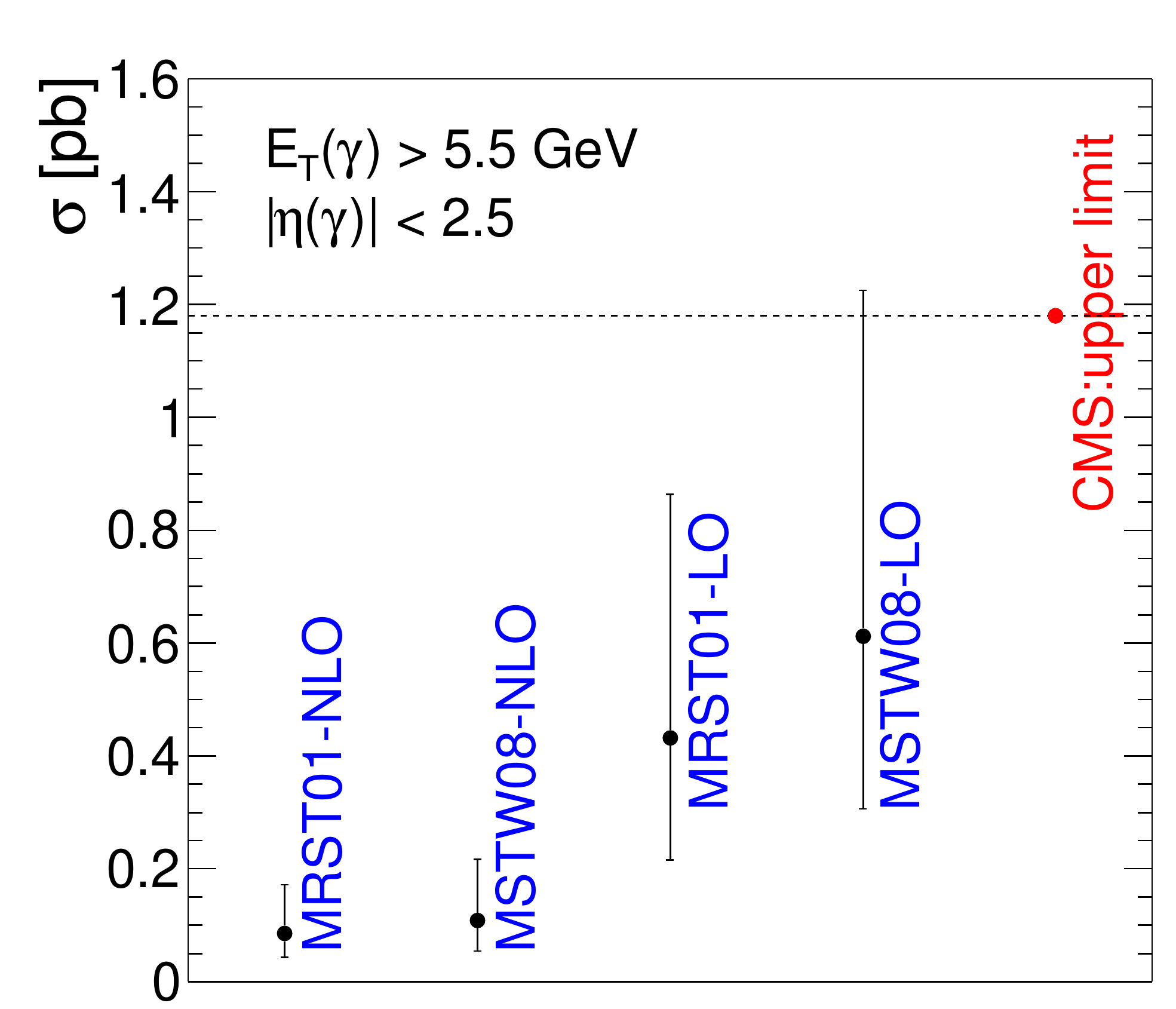}
\caption{Comparison of the upper limit (at 95\% CL) derived with the present data and four
theoretical predictions. The upper limit is on the sum of the exclusive and semi-exclusive
$\gamma\gamma$ production cross sections (where it is required that no particles from the proton 
dissociation have $|\eta| < 5.2$), while the theoretical predictions are for exclusive 
$\gamma\gamma$ production only. If the contributions from semi-exclusive production are included, 
the predictions increase by a factor of $\sim$2~\cite{private}. \label{fig:sigmacompare}}
\end{figure}

No diphoton events survive the selection criteria. An upper limit on the
production cross section is set employing a $CL_{s}$ approach~\cite{Junk:1999kv,0954-3899-28-10-313},
taking into account the
integrated luminosity, the selection efficiency, the
background contributions, and their uncertainties. A log-normal prior
is used for the integration over the nuisance parameters. This gives
an upper limit on the production cross section at 95\% confidence
level (CL):

\vspace{-1.5em}
\[\sigma(\ET(\gamma)>5.5\GeV,\,|\eta(\gamma)|<2.5)<1.18\unit{pb}.\]
The upper limit is on the sum of the exclusive (el-el) and semi-exclusive
(inel-el and inel-inel) $\gamma\gamma$ production cross sections, with no
particles from the proton dissociation having $|\eta| < 5.2$ for the semi-exclusive case.
Figure~\ref{fig:sigmacompare} shows the comparison between the present upper limit and the
predicted cross sections (el-el only) calculated with the \ExHuME generator.
Two different PDF sets,
MRST01~\cite{Martin:2001es,Martin:2002dr} and MSTW08~\cite{Martin:2009iq},
from both leading-order (LO) and next-to-leading-order (NLO) fits,
are considered. The difference between LO and NLO predictions
reflects mostly the difference in the low-$x$ gluon density.
The uncertainties in these theoretical predictions (in addition to those due to the PDFs) are
estimated to be a factor of about 2~\cite{private}, as shown in Fig.~\ref{fig:sigmacompare}. 
The upper limit measured in this analysis
is an order of magnitude above the predicted cross sections with
NLO PDFs, while it provides some constraint on the
predictions with LO PDFs. If the MSTW08-LO PDF is used,
the probability of finding no candidate in the present data is less than 23\%.
The semi-exclusive $\gamma\gamma$ production cross section has larger theoretical uncertainties, but is
expected to be of magnitude similar to that of the fully exclusive
process~\cite{private}.

\begin{figure}[hbtp]
\includegraphics[ width=0.5\textwidth]{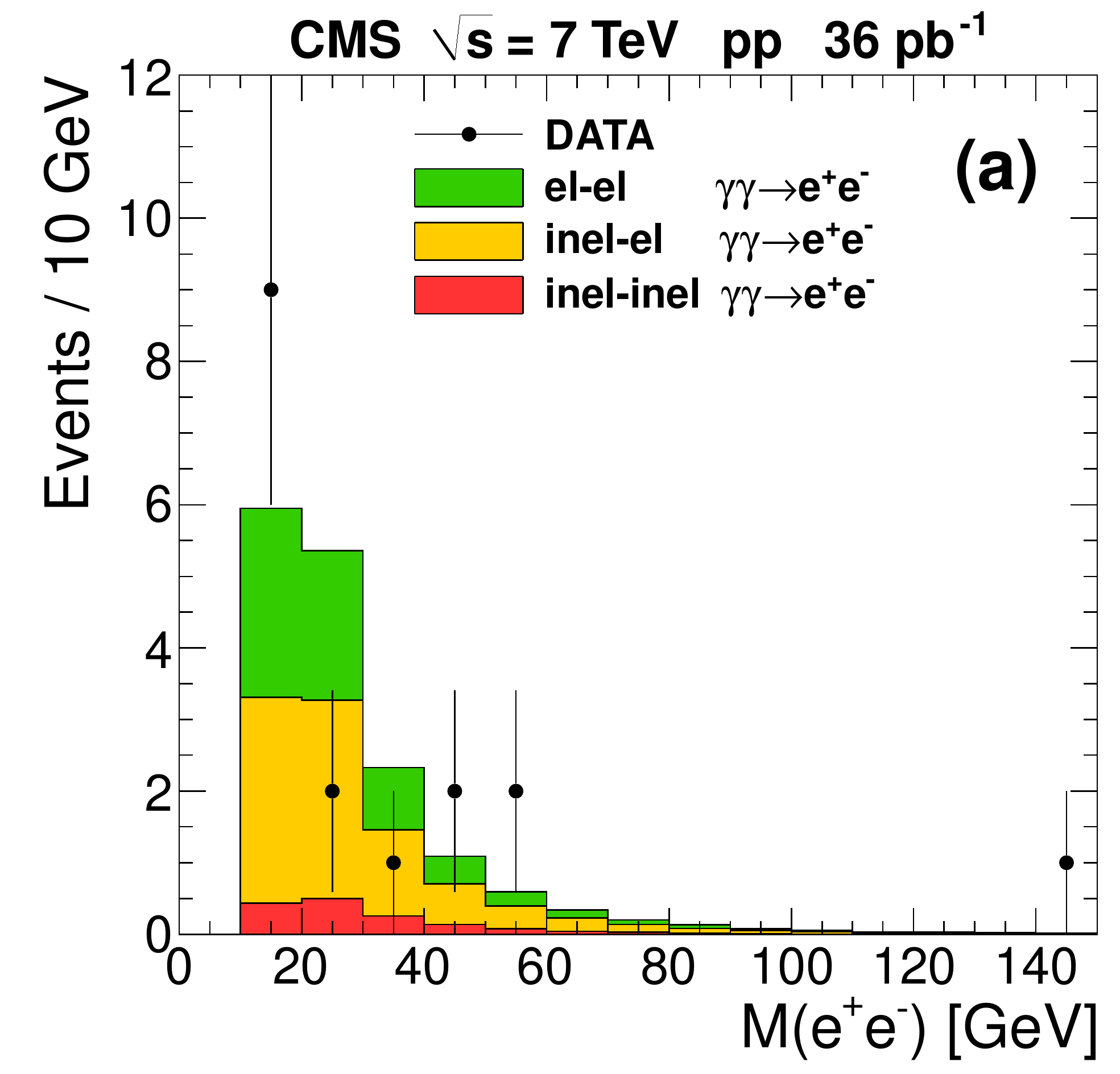}
\includegraphics[ width=0.5\textwidth ]{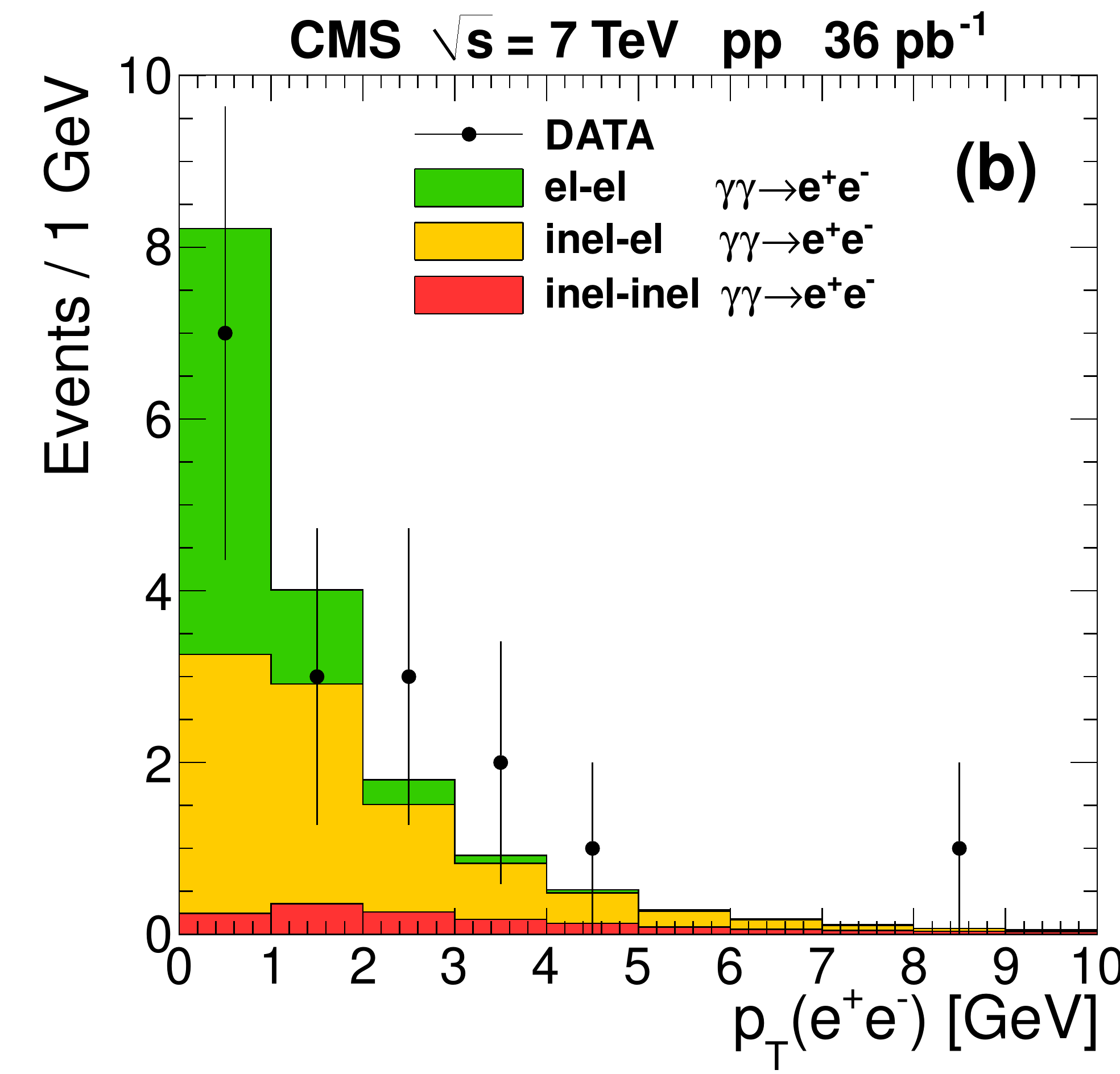}
\caption{Distributions of (a) the invariant mass and (b) the transverse momentum
of the \ee pairs, compared to the \LPAIR predictions (histograms) for the three
processes contributing to exclusive and semi-exclusive $\gamma
\gamma \rightarrow \ee$ production, passed through the full detector simulation and
reconstruction. The simulation is normalized to the integrated luminosity of the
data sample (36\pbinv), and does not include the estimated $0.85 \pm 0.28$ background events.
\label{fig:dielectronmcdatacomparedielectron} }
\end{figure}

\begin{figure}[hbtp]
\includegraphics[ width=0.5\textwidth]{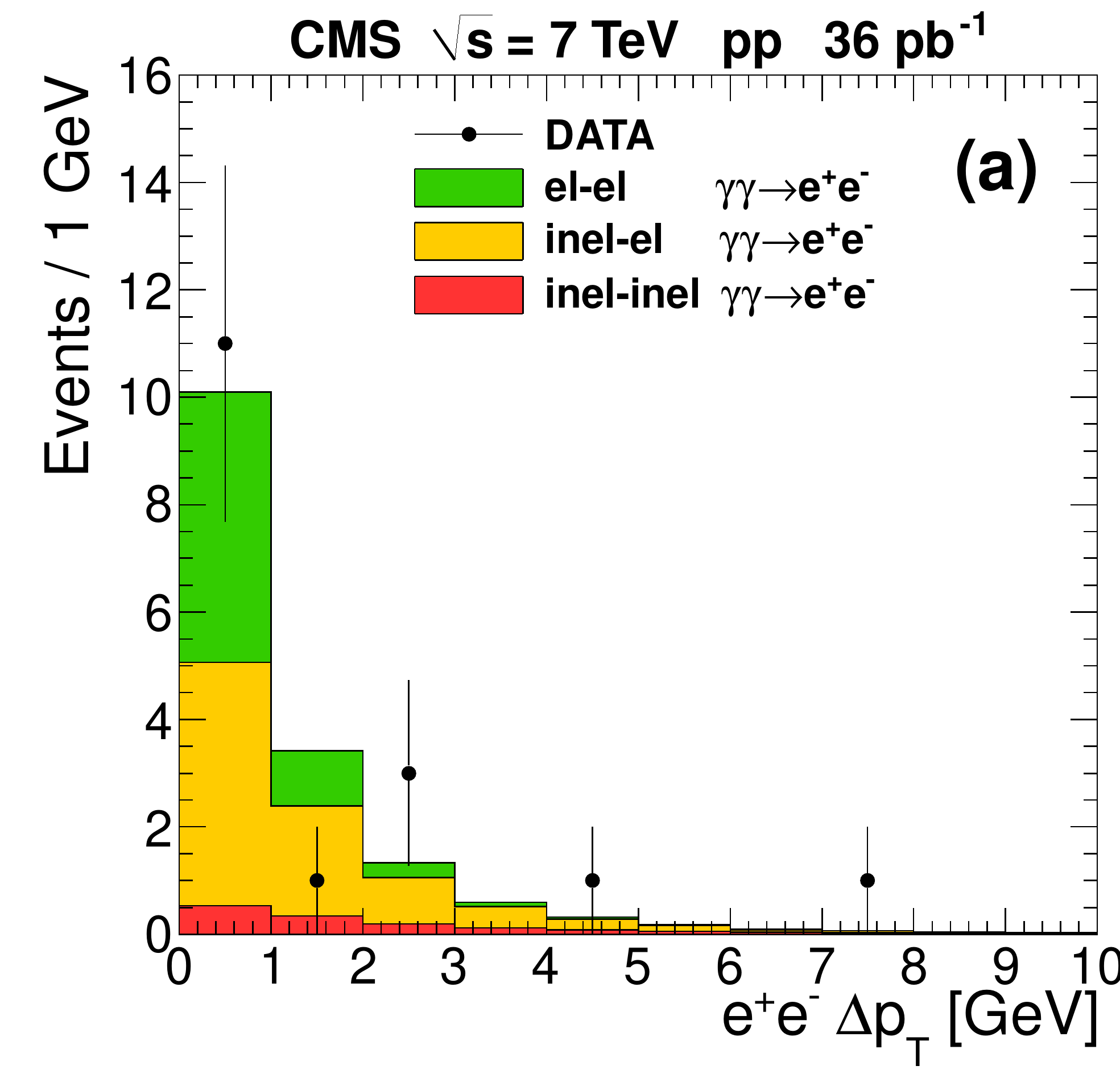}
\includegraphics[ width=0.5\textwidth ]{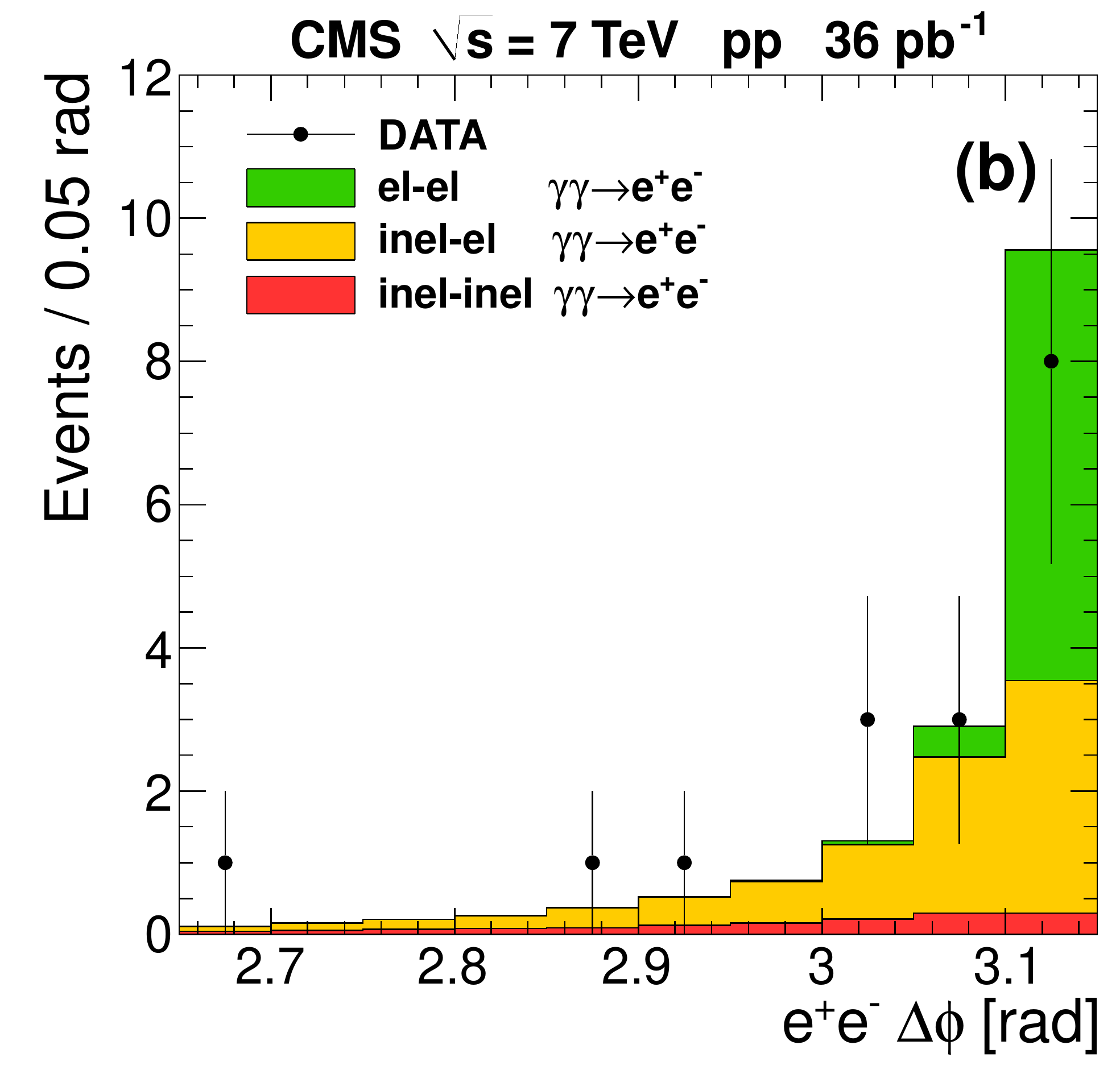}
\caption{Distributions of the difference of (a) the transverse momentum and (b) the
azimuthal angle of the \ee pairs, compared to the \LPAIR predictions (histograms)
for the three processes contributing to exclusive and semi-exclusive
$\gamma\gamma \rightarrow \ee$ production, passed through the full detector
simulation and reconstruction. The simulation is normalized to the integrated
luminosity of the data sample (36\pbinv), and does not include the estimated $0.85 \pm 0.28$ background events.
\label{fig:dielectronmcdatacomparedelta} }
\end{figure}

\begin{table}[hbtp]
\centering
\topcaption{Predicted \ee yields for
both exclusive and semi-exclusive \ee production. The relative uncertainty of the integrated luminosity \Lumi is 4\%~\cite{lumi}. The production cross sections $\sigma$ are calculated with
the \LPAIR generator. \label{tab:dielectronprediction}}%
\begin{tabular}{l|l|l|c|l}
\hline Process   & \Lumi (\pbinv)   & $\sigma$ (pb) & $\varepsilon$     & Yield (events)        \\
\hline el-el     &                  & 3.74          & 0.0481$\pm$0.0055 & 6.51$\pm$0.79\syst  \\
\cline{1-1} \cline{3-5}
       inel-el   & 36.2$\pm$1.4     & 6.68          & 0.0343$\pm$0.0042 & 8.29$\pm$1.07\syst  \\
\cline{1-1} \cline{3-5}
       inel-inel &                  & 3.52          & 0.0117$\pm$0.0019 & 1.49$\pm$0.25\syst  \\
\hline \hline Total &               &               &                   & 16.3$\pm$1.3\syst \\
\hline
\end{tabular}
\end{table}
Seventeen exclusive or semi-exclusive \ee candidates are observed, with an expected
background of $0.85\pm0.28\stat$ events, consistent with the
theoretical prediction for the combined el-el, inel-el and inel-inel
\ee yield of $16.3\pm1.3\syst$ events (Table~\ref{tab:dielectronprediction}).
Figure~\ref{fig:dielectronmcdatacomparedielectron} shows the comparison of the 
measured and simulated invariant-mass and \PT
distributions of the \ee pairs,
while Fig.~\ref{fig:dielectronmcdatacomparedelta} shows that for the  $\Delta\PT$
and $\Delta\phi$ distributions. Both the
yield and the kinematic distributions are consistent with the
assumption of exclusive and semi-exclusive \ee production via the $\gamma\gamma\rightarrow \ee$
process, which validates the analysis technique, notably the exclusivity
selection.


\section{Summary}
A search for exclusive or semi-exclusive $\gamma\gamma$ production and the
observation of exclusive and semi-exclusive \ee production have been presented, based on a
sample of pp collisions at $\sqrt{s}=7\TeV$ corresponding to an
integrated luminosity of 36\pbinv. Exclusive $\gamma\gamma$ production
helps improve the understanding of diffraction and provides a test of the
theoretical predictions for exclusive Higgs boson production. Exclusive
\ee production is dominantly a QED process and provides a means to check the
selection procedure for other exclusive processes. No diphoton events satisfy the selection criteria.
An upper limit on the cross section for the exclusive reaction
$\p\p \to \p+\gamma\gamma+\p$ and the corresponding semi-exclusive processes (in
which either or both protons diffractively dissociate and no particles from the proton
dissociation have $|\eta| < 5.2$), with $\ET(\gamma) >
5.5\GeV$ and $|\eta(\gamma)| < 2.5$, is set at 1.18 pb at 95\% confidence level. Using a similar
technique, 17 exclusive or semi-exclusive \ee candidates are observed, with an expected
background of $0.85\pm0.28\stat$ events, consistent with the \LPAIR
prediction of $16.3\pm1.3\syst$ events. Both the
number of candidates and the kinematic distributions are in
agreement with the expectation for exclusive and semi-exclusive \ee production via the
$\gamma\gamma\rightarrow \ee$ process.

\section*{Acknowledgements}
We congratulate our colleagues in the CERN accelerator departments for the excellent performance of the LHC machine. We thank the technical and administrative staff at CERN and other CMS institutes, and acknowledge support from BMWF and FWF (Austria); FNRS and FWO (Belgium); CNPq, CAPES, FAPERJ, and FAPESP (Brazil); MES (Bulgaria); CERN; CAS, MoST, and NSFC (China); COLCIENCIAS (Colombia); MSES (Croatia); RPF (Cyprus); MoER, SF0690030s09 and ERDF (Estonia); Academy of Finland, MEC, and HIP (Finland); CEA and CNRS/IN2P3 (France); BMBF, DFG, and HGF (Germany); GSRT (Greece); OTKA and NKTH (Hungary); DAE and DST (India); IPM (Iran); SFI (Ireland); INFN (Italy); NRF and WCU (Korea); LAS (Lithuania); CINVESTAV, CONACYT, SEP, and UASLP-FAI (Mexico); MSI (New Zealand); PAEC (Pakistan); MSHE and NSC (Poland); FCT (Portugal); JINR (Armenia, Belarus, Georgia, Ukraine, Uzbekistan); MON, RosAtom, RAS and RFBR (Russia); MSTD (Serbia); SEIDI and CPAN (Spain); Swiss Funding Agencies (Switzerland); NSC (Taipei); ThEP, IPST and NECTEC (Thailand); TUBITAK and TAEK (Turkey); NASU (Ukraine); STFC (United Kingdom); DOE and NSF (USA).

Individuals have received support from the Marie-Curie programme and the European Research Council (European Union); the Leventis Foundation; the A. P. Sloan Foundation; the Alexander von Humboldt Foundation; the Austrian Science Fund (FWF); the Belgian Federal Science Policy Office; the Fonds pour la Formation \`a la Recherche dans l'Industrie et dans l'Agriculture (FRIA-Belgium); the Agentschap voor Innovatie door Wetenschap en Technologie (IWT-Belgium); the Ministry of Education, Youth and Sports (MEYS) of Czech Republic; the Council of Science and Industrial Research, India; the Compagnia di San Paolo (Torino); and the HOMING PLUS programme of Foundation for Polish Science, cofinanced from European Union, Regional Development Fund.

\bibliography{auto_generated}
\cleardoublepage \appendix\section{The CMS Collaboration \label{app:collab}}\begin{sloppypar}\hyphenpenalty=5000\widowpenalty=500\clubpenalty=5000\textbf{Yerevan Physics Institute,  Yerevan,  Armenia}\\*[0pt]
S.~Chatrchyan, V.~Khachatryan, A.M.~Sirunyan, A.~Tumasyan
\vskip\cmsinstskip
\textbf{Institut f\"{u}r Hochenergiephysik der OeAW,  Wien,  Austria}\\*[0pt]
W.~Adam, T.~Bergauer, M.~Dragicevic, J.~Er\"{o}, C.~Fabjan\cmsAuthorMark{1}, M.~Friedl, R.~Fr\"{u}hwirth\cmsAuthorMark{1}, V.M.~Ghete, J.~Hammer, N.~H\"{o}rmann, J.~Hrubec, M.~Jeitler\cmsAuthorMark{1}, W.~Kiesenhofer, V.~Kn\"{u}nz, M.~Krammer\cmsAuthorMark{1}, D.~Liko, I.~Mikulec, M.~Pernicka$^{\textrm{\dag}}$, B.~Rahbaran, C.~Rohringer, H.~Rohringer, R.~Sch\"{o}fbeck, J.~Strauss, A.~Taurok, P.~Wagner, W.~Waltenberger, G.~Walzel, E.~Widl, C.-E.~Wulz\cmsAuthorMark{1}
% ...
\vskip\cmsinstskip
\dag:~Deceased\\
1:~~Also at Vienna University of Technology, Vienna, Austria\\
% ...

\end{sloppypar}
\end{document}